# Time–space height correlations of thermally fluctuating 2-d systems. Application to vicinal surfaces and analysis of STM images of Cu(1 1 5)

E. Le Goff, L. Barbier *, B. Salanon

*DSM/DRECAM/SPCSI, Bât 462, CEA/Saclay, F91191 Gif-sur-Yvette Cedex, France*

## Abstract

For thermally fluctuating 2-d systems, like solid surfaces, time and space correlation of the local surface height diverge logarithmically in the rough phase, whereas saturation is obtained below the roughening transition. A 2-d Langevin formalism, allowing to recover for long times and/or large distances these asymptotic behaviors, is presented. An overall expression for correlation functions that are related to atom hopping rates and surface stiffnesses is given. Considering anisotropic systems allows describing vicinal surfaces. At finite times, time correlations cross over to power laws $\alpha t^{1/n}$ ($n = 1$, 2 or 4), within limited time ranges as it was observed for isolated fluctuating steps. Limits of time ranges are related to stiffnesses and diffusion anisotropies. Application to the analysis of STM images of Cu(1 1 5) below the roughening transition is given.
© 2003 Elsevier Science B.V. All rights reserved.

## 1. Introduction

Today scanning tunneling microscopy (STM) routinely provides images of surfaces at the atomic scale. Valuable informations on the local structure usually results. For fluctuating surfaces, the equilibrium properties (energetic) and the dynamics of the system can be also investigated. For this purpose, the statistical analysis of a set of STM images is the necessary tool for extracting relevant information such as distribution of individual objects or correlation functions. Results of the measurement of such quantities have further to be analyzed in terms of the elementary properties of the surface at the atomic scale. In that way, wide efforts were made to interpret STM images of vicinal surfaces at thermal equilibrium. It was mainly shown that spatial correlation functions of step edge positions can be related to the elementary kink energy and to the step–step interaction [1], whereas time correlation functions allows identifying atomic diffusion processes and their associated activation energies [2]. Knowledge of such quantities is of great interest for understanding and predicting the stability of these surfaces, which determines the shape of nanocrystals. Applications to the stability of nanostructures follow together with a better understanding of their physical and reactivity properties.

* Corresponding author. Tel.: +33-1-6908-5160; fax: +33-1-6908-8446.
*E-mail address:* lbarbier@cea.fr (L. Barbier).

Observations of fluctuating systems are the source of valuable information on their dynamics. Random forces (noise) allow the system to fluctuate locally whereas matter conservation and the energetic of the whole system acting as a thermostat contain the fluctuations. These are the basis of phenomenological Langevin dynamic equations. As for surfaces, 2-d Langevin equation (where surface tensions and diffusion constants are enough to characterize the system) are a useful tool allowing the analysis of STM measured correlation functions.

The time correlation function $G(t)$ ($G(t) = \langle (h(y) - h(0))^2 \rangle$, $h(y)$: step displacement at ordinate $y$) of one thermally fluctuating step has been observed to vary as $G(t) = \alpha t^{1/n}$, $n = 2$ or 4. Considering here fluctuations of one single step, 1-d approaches have shown that the exponent value depends on the atomic diffusion process: at low temperature ($T$), where atoms emitted from kink sites at step edges preferentially diffuse along step edges, a $t^{1/4}$ dependence is expected, whereas at higher temperature, where steps fluctuate under attachment or detachment of terrace adatoms $G(t)$ follows a $t^{1/2}$ power law [3–8]. Intermediate time dependence ($n = 3$) has been also predicted (although not observed for metal surfaces in UHV [2], but observed at liquid–solid interfaces [9]) for slow attachment detachment of atoms from one step edge and fast diffusion within the adjacent terrace [3,4,10]. An exhaustive review of that field can be found in [2].

Step–step interactions are responsible for the stability of vicinal surfaces and although they are very weak, their influence on step fluctuations at thermal equilibrium cannot be neglected. Similarly to the collision length associated to spatial fluctuation $y_c \approx L^2 \eta_y / 4kT$ [11], where $L$ is the average step–step distance and $\eta_y$ the step stiffness, a collision time can be defined as the time for an isolated step to fluctuate over a distance $L/2$: $t_c \approx (L^2/4\alpha)^n$, $n = 1$–4 (here and in the following $x$ (resp. $y$) designates the direction perpendicular (resp. parallel) to step edges. The isolated step approximation can be considered for $t \ll t_c$, whereas for $t > t_c$, step–step interactions are expected to reduce step fluctuations and make the isolated step approximation no longer valid. Step–step interactions are both entropic (step cannot cross) and elastic with the usual $A/L^2$ dependence, where $A$ is the interaction strength). These two interactions do not add-up linearly and an analytical expression for the total free energy interaction was obtained from the analytical solution of the 1-d interacting fermion model by Sutherland [12] (see also [13,14]). The step-fermion mapping leads to useful analytical expressions for the surface stiffnesses $\eta_x$ and $\eta_y$ [15] in terms of the two energetic parameters: the elementary kink energy $E_k$ and $A$. Measurements of the spatial correlation of step displacements allow to get the energetic parameters of a vicinal surface, namely the kink creation energy $E_k$ and the elastic step–step interaction constant $A$ [1,16,17].

Beyond the ideal case of isolated steps, various attempts have been made to analyze time fluctuations of interacting steps on vicinal surfaces. A Langevin formalism, for a single step in an external harmonic potential was first considered [3,7,18,19]. Further attempts have been made to take into account the step–step interaction by a hard wall approximation [20] or by considering the fluctuations of an individual step within a step array [21]. Correlated mass transport across terraces was also found to induce strong time correlations between adjacent steps (even for non-interacting steps) [22]. Moreover, it has been noticed that $G(t)$ must ultimately diverges logarithmically like the static roughness [10,21,23]. However, for strongly interacting steps and particularly for a vicinal surface below the Kosterlitz-Thouless roughening transition these approaches cannot be used and a full 2-d analysis is required.

In the present paper, we report on a 2-d Langevin formalism for an anisotropic thermally fluctuating surface above and below its roughening transition temperature ($T_R$). In contrast with previous works dealing with fluctuations of a rough interface in equilibrium with a vapor (non-conserved order parameter) [23], atom diffusion restricted within the surface plane (conserved order parameter) is here considered. Our model is a 2-d extension of 1-d Langevin approaches [2–6], and of the capillary wave model, that provides useful expressions for space correlation functions above and below $T_R$ [1]. Above $T_R$, the spatial correlation functions of surface heights diverges logarithmically and our approach confirms that the time correlation function diverges like $\ln(t)$ as well. It is here emphasized that for all rough vicinal surfaces, whose time correlation functions can follow a power law behavior within limited time ranges, step–step interactions ultimately reduce the time variation to a logarithmic divergence for long times. For short times, power law behaviors are recovered, which exponents $1/n$ are found depending on the time range and on the energetic and diffusional anisotropies. In addition, saturation of $G(t)$ below $T_R$ is obtained as expected. Experimental results on the Cu(1 1 5) vicinal surface are found to be in agreement with this prediction.

## 2. 2-d Langevin formalism

For a full description of space–time fluctuations of a solid surface, a 2-d approach is mandatory. We have thus developed a generic 2-d Langevin formalism for a thermally fluctuating 2-d system. Keeping in mind applications to vicinal surfaces, the diffusional and energetic anisotropies are carefully taken into account. Furthermore surfaces above ($T > T_R$) as well as below ($T < T_R$) their roughening transition are here considered. To keep an easy link with previous 1-d descriptions, appropriate notations for application to vicinal surfaces are used. One step configuration of a vicinal surface can be described by the set of displacements of step number $m$ at ordinate $y$: $\{h_{m,y}\}$ (to be more generic, the notation $h(x, y)$ could be also used for local surface heights and $h(x(m), y)$ for a vicinal surface). For the ideally ordered surface, $x(m) = mL$, $L$ being the nominal step–step distance. Fig. 1 sum up the notations used throughout the text.

On a mesoscopic scale, a vicinal surface can be seen as an anisotropic fluctuating surface. We use the following model able to describe its energetic:

Within the capillary wave model, the Hamiltonian $H$ of an array of steps can be written [24,25]:

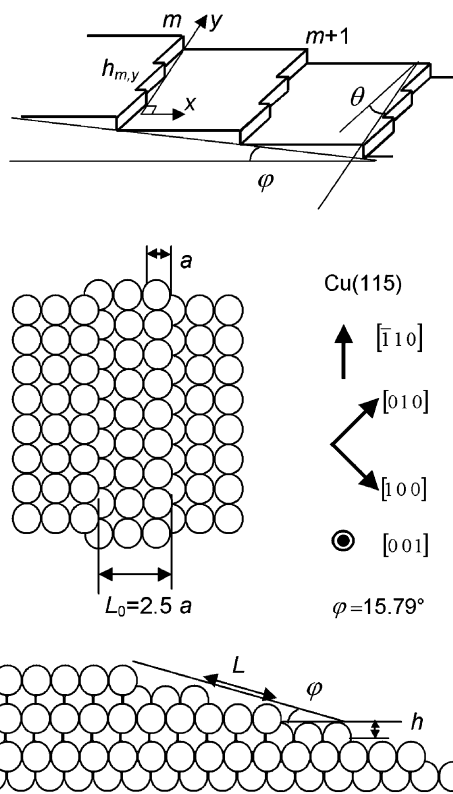

Fig. 1. Notations used throughout the text and top view and cross-section of Cu(1 1 5) where [1 1 0] is the step down direction. $\varphi$ is the miscut angle (angle between terrace and surface planes), $\theta$ is the step orientation with respect to one dense step direction. $x$ and $y$ are perpendicular axis ($y$ along step edges).

$$H = \sum_{m,y} \left( \frac{\eta_x}{2} (h_{m+1,y} - h_{m,y})^2 \right.$$
$$\left. + \frac{\eta_y}{2} (h_{m,y+1} - h_{m,y})^2 + V(h_{m,y}) \right) \quad (1)$$

or equivalently in the Fourier space (see Appendix A):

$$H = \sum_{q_x, q_y} [\eta_x (1 - \cos(qx))^2 + \eta_y (1 - \cos(qy)) + 4\pi^2 V_{\mathrm{loc}}]|h_q^2|. \quad (2)$$

In the above expressions, the localization potential $V(h_m(y))$ favors integer values of the local surface displacements $h_m(y)$, which allows describing a surface below its roughening transition ($T < T_R$). $\eta_x$ and $\eta_y$ are the surface stiffnesses that depend on temperature (as shown in [1], $\eta_x$ and $\eta_y$, for a vicinal surface at $T > T_R$, are simply related to the microscopic parameters $E_k$ and $A$). However, note that analytical expressions for the localization potential $V_{loc}$ and the surface stiffnesses below $T_R$ are still not known.

The very 1-d formalism used for describing isolated steps [3,7] or one step within an array [21], can be extended for describing 2-d fluctuations. In the Langevin formalism, the surface is assumed to relax locally to minimize the system energy. The back force depends on the diffusion processes. Without matter conservation (i.e. for evaporation–condensation of atoms at one isolated step), this force is simply proportional to the energy gradient and the 1-d Langevin equation is in that case [3]:

$$\frac{\partial h(y)}{\partial t} = -\frac{\Gamma}{kT}\left(\frac{\delta H}{\delta h(y)}\right) + \xi(y,t), \tag{3}$$

where $h(y)$ is the local step displacement, $\Gamma$ is an atom hopping rate (also sometimes called friction coefficient) and $\xi(y,t)$ is a noise term allowing thermal fluctuations. This noise is uncorrelated in space and time and its correlation function can be written:

$$\langle \xi(y',t')\xi(y,t) \rangle = 2\Gamma \delta(y-y')\delta(t-t'). \tag{4}$$

Note that the very same constant $\Gamma$ appears in Eqs. (3) and (4), making the principle of energy equipartition at thermal equilibrium satisfied.

For atom diffusion restrained along step edges, return at the equilibrium position is slowed down by local matter conservation. In that case, the constant $-\Gamma$ within (3) has to be replaced by $+\Gamma(\partial^2/\partial x^2)$ [3] making the back force depending on the local step curvature. The noise correlation function is accordingly modified to reflect matter conservation:

$$\langle \xi(y',t')\xi(y,t) \rangle = -2\Gamma\left[\frac{\partial^2}{\partial y^2}\delta(y-y')\right]\delta(t-t'). \tag{5}$$

We now introduce a 2-d anisotropic Langevin equation assuming local matter conservation for diffusion in both directions (for a vicinal surface, atoms are a priori allowed to jump along step edges as well as between step edges). To perform this calculation, we assume that surface diffusion can be described by means of two effective hopping rates $\Gamma_x$, $\Gamma_y$ (respectively, for a vicinal surface, in the perpendicular and parallel to step edges direction), containing all details of the diffusion processes. Hopping rates has further to be related to diffusion coefficients of elementary diffusion processes.

For a locally matter conserving system, a 2-d Langevin equation can be written:

$$\frac{\partial h_m(y)}{\partial t} = \frac{1}{kT}\left(\Gamma_x\frac{\partial^2}{\partial m^2} + \Gamma_y\frac{\partial^2}{\partial y^2}\right)\left(\frac{\delta H}{\delta h_m(y)}\right) + \xi(m,y,t). \tag{6}$$

Within the above expression

$$\Delta_a = \left(\Gamma_x\frac{\partial^2}{\partial m^2} + \Gamma_y\frac{\partial^2}{\partial y^2}\right)$$

is an anisotropic Laplacian reflecting anisotropic local diffusion processes (making the back force depending on local surface curvature). Application of Eq. (6) is limited for time much larger than the shortest hopping times ($1/\Gamma_x$ or $1/\Gamma_y$). In order to satisfy the matter conservation rule and the energy equipartition principle (in a similar way as for Eqs. (4) and (5)), the correlation function of the noise term $\eta(x,y,t)$ must be written:

$$\langle \xi(m',y',t')\xi(m,y,t) \rangle$$
$$= -2\left(\Gamma_x\frac{\partial^2}{\partial m^2}\delta(m-m')\right.$$
$$\left.+ \Gamma_y\frac{\partial^2}{\partial y^2}\delta(y-y')\right)\delta(t-t'). \tag{7}$$

Solving Eq. (6) allows writing the space–time correlation function of step fluctuations at thermal equilibrium:

$$G(m,y,t) = \langle (h_m(y,t) - h_0(0,0))^2 \rangle \tag{8}$$

as a sum in the Fourier space (for calculation details see Appendix A):

$$G(m,y,t) = \frac{kT}{N_xN_y}\sum_{q_x,q_y}\frac{1-e^{-a_q|t|}e^{-i(q_xm+q_yy)}}{b_q}, \tag{9}$$

where

$$a_q = \frac{4b_q}{kT}[\Gamma_x(1-\cos(q_x)) + \Gamma_y(1-\cos(q_y))]$$

with

$$b_q = \eta_x(1-\cos(q_x)) + \eta_y(1-\cos(q_y)) + 4\pi^2 V_{loc}. \tag{10}$$

The time correlation function is:

$$G(0,0,t) = G(t) = \langle (h_m(y,t) - h_m(y,0))^2 \rangle$$
$$= \frac{kT}{N_x N_y} \sum_{q_x,q_y} \frac{1 - e^{-a_q|t|}}{b_q}. \quad (11)$$

Within Eq. (9), $m$ and $y$ are spatial coordinates (step edge positions for a vicinal surface). On usual STM images, where the STM tip scans the surface, measurement of step positions distant by $(m,y)$ are separated in time by $t$ and one measures $G(m,y,t(m,y))$ where $t$, $m$ and $y$ are related by the scanning speed. Within Eq. (11), $G(t)$ is the time correlation function for the height of one surface site ($m$ and $y$ being fixed and taken as origin $((m,y) = (0,0))$). In STM studies, such measurement is obtained for example by scanning the tip repetitively along $x$ at one fixed ordinate $y$.

For $t = 0$ within Eq. (9), well known expressions for spatial correlation functions are recovered [1,15]:

$$G(m,y) = \frac{kT}{N_x N_y}$$
$$\times \sum_{q_x,q_y} \frac{1 - \cos(q_x m + q_y y)}{\eta_x(1 - \cos(q_x)) + \eta_y(1 - \cos(q_y)) + 4\pi^2 V_{\text{loc}}}. \quad (12)$$

Note that single integral and asymptotic analytical forms were given for $G(m,0)$ and $G(0,y)$ in [1,15].

Eqs. (9), (11) and (12) are very general and valid for every anisotropic 2-d fluctuating system. Application of these relations only needs knowledge of surface stiffnesses and hopping rates.

## 3. Discussion

### 3.1. Isotropic system

Eq. (9), or its integral form in a continuous approximation, can be easily computed. In the following, we would emphasize some properties of the time correlation function $G(t)$ (Eq. (11)). One considers first an isotropic system ($\Gamma = \Gamma_x = \Gamma_y$ and $\eta = \eta_x = \eta_y$) at thermal equilibrium above its roughening transition ($V_{\text{loc}} = 0$). For very short times $t \ll (1/4\Gamma)(kT/\eta)$ (i.e. in conditions out of the hypothesis of the Langevin approach), one gets:

$$G(t) = 8\Gamma t. \quad (13)$$

For larger times $G(t)$ behaves as:

$$G(t) \approx \frac{kT}{\pi} \frac{1}{\eta} \int_0^\pi \frac{1 - \exp\left(-\frac{\Gamma}{kT}\eta u^4 |t|\right)}{u} \, du. \quad (14)$$

And for long times ($t \gg kT/\Gamma\eta$), $G(t)$ reduces to:

$$G(t) \approx \frac{kT}{\pi\eta}\left(\frac{1}{4}\ln\left(\frac{\pi^4 \Gamma\eta}{kT}t\right) + C\right), \quad (15)$$

where a logarithmic divergence of the correlation function is obtained. The additive constant $C$ is numerically found equal to 0.7326. The logarithmic divergence is characteristic of fluctuations of a two dimensional system. Similarly, we find that the spatial correlation function diverges as $(kT/\pi\eta) \times (\ln(r) + 2)$ [15] where $r$ is a spatial coordinate ($m$ or $y$). Considering here a system in its rough phase, the logarithm prefactor must be larger than the universal value $2/\pi^2$ and the above relations apply for $kT/\eta > 8/\pi$. Fig. 2 shows the behavior of $G(t)$ in reduced coordinates. One sees that for an isotropic surface the logarithmic behavior is already reached at short times and that power law behaviors with $1/n$ exponents ($n = 4$ or 2) are not present in that case.

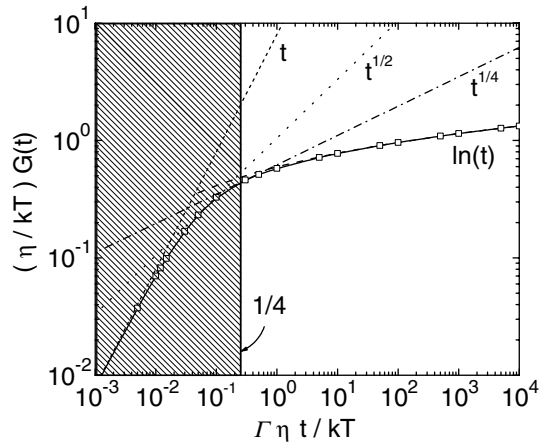

Fig. 2. (-□-) $G(t)$ (Eq. (9)) for a rough ($V_{\text{loc}} = 0$ K) isotropic surface. The various dashed and dotted lines indicates linear, $t^{1/4}$, $t^{1/2}$ and $\ln(t)$ behaviors for comparison. The hatched area ($t < kT/4\Gamma\eta$) is a time region where the Langevin approach is irrelevant.

Below $T_R$ ($kT/\eta < 8/\pi$), $V_{loc}$ becomes positive and the very presence of the $4\pi^2 V_{loc}$ term within $b_q$ makes temporal and spatial correlation function saturate for long times and large distances to the same value: For $V_{loc} < \eta/8\pi^2$ (i.e. slightly below $T_R$) saturation occurs after a quasi-logarithmic regime and one gets:

$$G(t \to \infty) = G(r \to \infty) \approx \frac{kT}{\eta\sqrt{2 + 8\pi^2 \frac{V_{loc}}{\eta}}}, \quad (16)$$

whereas for $V_{loc} > \eta$, the correlation function is in the linear regime before saturation to:

$$G(t \to \infty) = G(r \to \infty) \approx \frac{kT}{4\pi^2 V_{loc}}. \quad (17)$$

### 3.2. Anisotropic system, $T > T_R$

We consider now a rough ($V_{loc} = 0$) strongly anisotropic system where (as for usual vicinal surfaces) $\eta_x \ll \eta_y$ and $\Gamma_x \ll \Gamma_y$. Depending on the dominant term within $a_q$, (and thence on the considered time range) different analytical expressions can be obtained as shown in Appendix A and as summarized in Table 1.

Similarly to the isotropic case, for very short times ($t \ll t_1 = kT/4\Gamma_y\eta_y$, i.e. beyond the strict validity of the Langevin description) one obtains the $\eta_x$ and $\Gamma_x$ independent expression:

$$G(t) \approx 4\Gamma_y t. \quad (18)$$

For short times, and provided:

$$t_1 < t < t_2 = \frac{kT}{4\Gamma_y\eta_y}\left(\frac{\eta_x}{\eta_y} + \frac{\Gamma_x}{\Gamma_y}\right)^{-2}$$

Table 1
Depending on time and the stiffness and hopping rate anisotropies, simple analytic forms are obtained for the time correlation function from the very same basic equation (Eq. (11))

|  | Time interval | Time correlation function |
|---|---|---|
| *Isotropic case* | | |
| | $t < \frac{1}{4\Gamma}\frac{kT}{\eta}$ | $G(t) = 8\Gamma t$ |
| | $t > \frac{1}{\Gamma}\frac{kT}{\eta}$ | $G(t) \approx \frac{T}{\pi\eta}\left(\frac{1}{4}\ln\left(\frac{\pi^4\Gamma\eta}{T}t\right) + \text{cte}\right)$ |
| *Anisotropic case* | | |
| | $t \ll \frac{kT}{4\Gamma_y\eta_y}$ | $4\Gamma_y t$ |
| | $\frac{kT}{4\Gamma_y\eta_y} < t < \frac{kT}{4\Gamma_y\eta_y}\left(\frac{\eta_x}{\eta_y} + \frac{\Gamma_x}{\Gamma_y}\right)^{-2}$ | $2\frac{\Gamma(3/4)}{\pi}\left(\frac{kT}{\eta_y}\right)^{3/4}(\Gamma_y t)^{1/4}$ |
| $\frac{\eta_y}{\eta_x} \gg \frac{\Gamma_y}{\Gamma_x}$ | $\frac{kT}{4\Gamma_y\eta_y}\left(\frac{\eta_x}{\eta_y} + \frac{\Gamma_x}{\Gamma_y}\right)^{-2} < t < \frac{kT}{4\Gamma_y\eta_x}\left(\frac{\eta_x}{\eta_y} + \frac{\Gamma_x}{\Gamma_y}\right)^{-1}$ | $8\pi^{-3/2}\left[\frac{kT}{\eta_y}\left(\frac{\eta_x}{\eta_y} + \frac{\Gamma_x}{\Gamma_y}\right)\Gamma_y t\right]^{1/2}$ |
| | $\frac{kT}{4\Gamma_y\eta_x}\left(\frac{\eta_x}{\eta_y} + \frac{\Gamma_x}{\Gamma_y}\right)^{-1} < t$ | $\frac{kT}{\pi\sqrt{\eta_x\eta_y}}\left[\frac{1}{4}\ln\left(\pi^4\frac{\eta_x}{kT}\Gamma_x t\right) + C\right]$ |
| $\frac{\eta_y}{\eta_x} \ll \frac{\Gamma_y}{\Gamma_x}$ | $\frac{kT\eta_y}{4\Gamma_y\eta_x^2} < t$ | $\frac{kT}{\pi\sqrt{\eta_x\eta_y}}\left[\frac{1}{4}\ln\left(\pi^4\frac{\eta_x^2}{\eta_y kT}\Gamma_y t\right) + C\right]$ |

one gets the following expression that is also $\eta_x$ and $\Gamma_x$ independent:

$$G(t) = 2\frac{\Gamma(3/4)}{\pi}\left(\frac{kT}{\eta_y}\right)^{3/4}(\Gamma_y t)^{1/4}. \quad (19)$$

In this time interval, diffusion along the $y$ axis dominate the system. The usual expression for time fluctuations of isolated steps with diffusion restricted to step edges [7] is recovered. In this $t$ range, surface stiffnesses and/or diffusion anisotropies make the system look one dimensional.

For longer times, such that

$$t_2 < t < t_3 = \frac{kT}{4\Gamma_y \eta_x}\left(\frac{\eta_x}{\eta_y} + \frac{\Gamma_x}{\Gamma_y}\right)^{-1}$$

one gets:

$$G(t) = 8\pi^{-3/2}\left[\frac{kT}{\eta_y}\left(\frac{\eta_x}{\eta_y} + \frac{\Gamma_x}{\Gamma_y}\right)\Gamma_y t\right]^{1/2}. \quad (20)$$

In this regime, step fluctuations are correlated. Assuming first that the energetic anisotropy is much higher than the diffusion one ($\eta_y/\eta_x \gg \Gamma_y/\Gamma_x$), one gets:

$$G(t) = 8\pi^{-3/2}\left[\frac{kT}{\eta_y}\Gamma_x t\right]^{1/2}, \quad (21)$$

where atom diffusion along the $x$ axis (between neighboring steps for a vicinal surface) induces the correlation. For very long times $t > t_3$, one gets:

$$G(t) = \frac{kT}{\pi\sqrt{\eta_x\eta_y}}\left[\frac{1}{4}\ln\left(\pi^4\frac{\eta_x}{T}\Gamma_x t\right) + C\right], \quad (22)$$

where the added constant $C$ is equal to that of the isotropic case. A logarithmic divergence of the time correlation function is obtained. This is reminiscent of spatial correlation functions for an anisotropic system that diverge similarly according to:

$$G(m,0,0) = \frac{kT}{\pi\sqrt{\eta_x\eta_y}}[\ln(m) + 2] \quad \text{and}$$
$$G(0,y,0) = \frac{kT}{\pi\sqrt{\eta_x\eta_y}}\left[\ln(y) + \frac{3}{2}\right] \quad [15]. \quad (23)$$

Thus, $G(t)$ may vary like to $t^{1/n}$ ($n = 1$, 2 or 4) or like to $\ln(t)$, depending on the time range and on the stiffness/hopping rate anisotropies. This is illustrated in Fig. 3 where parameters are chosen in order to produce well distinct regimes in the rough phase.

Assuming now that the system anisotropy is governed by the diffusion ($1 \ll \eta_y/\eta_x \ll \Gamma_y/\Gamma_x$), the extended $t^{1/2}$ regime shrinks in that case, as upper limits for the $t^{1/2}$ regime and the $t^{1/4}$ regime become similar ($t_2 \approx t_3$). For long times $t > t_3 = t_2$, one has:

$$G(t) = \frac{kT}{\pi\sqrt{\eta_x\eta_y}}\left[\frac{1}{4}\ln\left(\pi^4\frac{\eta_x^2}{\eta_y T}\Gamma_y t\right) + C\right], \quad (24)$$

where $C$ is again the same as for the isotropic case. Fig. 4 illustrate the $G(t)$ behavior in that case.

In contrast with previous 1-d approaches of step dynamics, the above expressions show that the time correlation function depends on both energetic and hopping rates in the two directions. According to the expression of $t_1$, $t_2$ and $t_3$ intermediate $t^{1/4}$ or $t^{1/2}$ regimes after the linear and before the logarithmic regimes only exist for systems with some energetic anisotropy (whatever is the diffusion anisotropy). The diffusion anisotropy governs the relative extension of the $t^{1/4}$ (strong diffusion anisotropy) and $t^{1/2}$ regimes (low diffusion anisotropy) in the intermediate time range ($t_1 < t < t_3$) (see also Section 4). As already pointed out in [20] and experimentally observed for vicinals of Cu(1 1 1) around 500 K and for Ag(1 1 1) around 450 K [2], a gradual transition from $t^{1/4}$ to $t^{1/2}$ may occur. This is the signature that these systems are governed by the step stiffness anisotropy ($\eta_y \gg \eta_x \Gamma_y/\Gamma_x$).

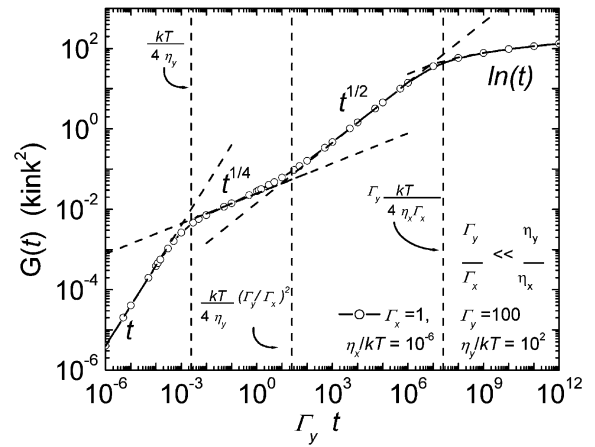

Fig. 3. (-○-) $G(t)$ (as obtained from Eq. (11)) for a rough ($V_{\text{loc}} = 0$ K) anisotropic surface ($\Gamma_x/\Gamma_y \gg \eta_x/\eta_y$). Parameters are chosen to get well-separated regimes. The various dashed and dotted lines indicates the $t$ (Eq. (18)), $t^{1/4}$ (Eq. (19)), $t^{1/2}$ (Eq. (20)) and $\ln(t)$ (Eq. (22)) approximations according to Table 1 and vertical dashed lines gives the limit between the various regimes.

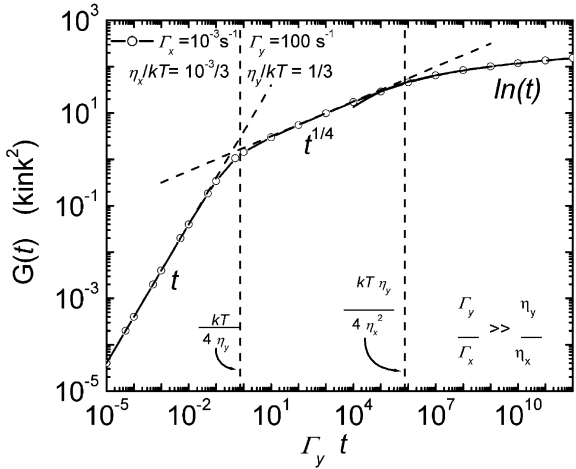

Fig. 4. (-○-) $G(t)$ (as obtained from Eq. (11)) for a rough ($V_{loc} = 0$ K) anisotropic surface ($\eta_x/\eta_y \gg \Gamma_x/\Gamma_y$). Parameters are chosen to get well-separated regimes. The various dashed and dotted lines indicates the $t$ (Eq. (18)), $t^{1/4}$ (Eq. (19)), and $\ln(t)$ (Eq. (24)) approximations according to Table 1. An extended $t^{1/2}$ (Eq. (20)) regime is not present in that case.

### 3.3. $T < T_R$

Below $T_R$, both temporal and spatial correlation functions saturate to the very same value. The localization potential $V_{loc}$ is not zero and is an increasing function of $(T_R - T)$. For $\eta_x \ll \eta_y$ and $V_{loc} \gg \eta_x$, one gets:

$$G(t \to \infty) = G(r \to \infty)$$
$$\approx \frac{kT}{\sqrt{2\eta_y(\eta_x + 4\pi^2 V_{loc})}}. \quad (25)$$

For $8\pi^2(V_{loc}/\eta_y) > 1 - 2(\eta_x/\eta_y)$, $G(t)$ saturates to this value after the linear regime. For temperature closer to $T_R$, $V_{loc}$ decreases and saturation occurs within the $t^{1/4}$ regime provided one has:

$$8\pi^2 \frac{V_{loc}}{\eta_x} > \frac{1}{2}\left(\left(\frac{\pi}{\Gamma(3/4)}\right)^2 \left(1 + \frac{\Gamma_x \eta_y}{\Gamma_y \eta_x}\right) - 4\Gamma(3/4)^2\right). \quad (26)$$

For lower $V_{loc}$ values ($T$ closer to $T_R$) the $t^{1/2}$ regime (for $\eta_y/\eta_x \gg \Gamma_y/\Gamma_x$) is reached prior to saturation and it is only in a narrow temperature region below $T_R$ that saturation may occur after a transient logarithmic regime. To illustrate these results for $T < T_R$, $G(t)$ is shown on Fig. 5 for various $V_{loc}$ values (as for comparison, typical values of $V_{loc}$ for Cu(1 1 5) within the 300–365 K temperature range are given in Table 4, much higher values can be expected for $T < 293$ K), while keeping constant the other parameters ($\Gamma_x$, $\Gamma_y$, $\eta_x$ and $\eta_y$).

## 4. Vicinal surfaces

To illustrate the results of our 2-d Langevin formalism, direct application to vicinal surfaces is now considered. Energetic parameters and diffusion properties were extensively measured and/or predicted by theoretical calculations for vicinals of Cu(1 1 1) or Cu(0 0 1), giving a set of data for these surfaces that can be used to illustrate on a realistic system some results of our approach. In that way, using reasonable values of the parameters, the temperature interval and the terrace width range where the diffusion anisotropy dominates can be given ($T$–$L$ diagram). Also given, ($t$–$T$) and ($t$–$L$) diagrams showing, as a function of temperature or terrace width, the extension of the various regimes for $G(t)$. These results are displayed from Figs. 7–9.

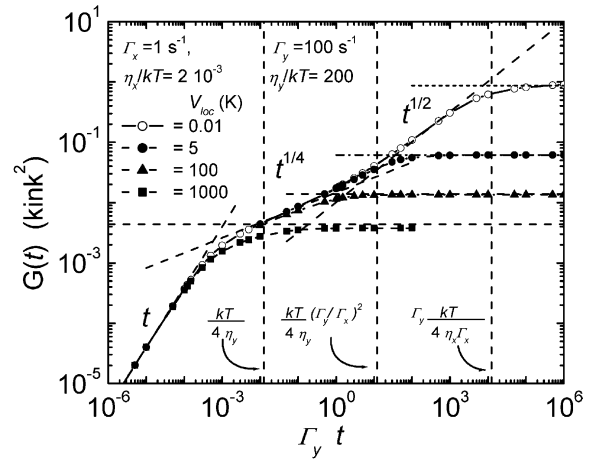

Fig. 5. $G(t)$ (as obtained from Eq. (11)) for a flat ($V_{loc} > 0$) anisotropic surface ($\eta_x/\eta_y \ll \Gamma_x/\Gamma_y$). Depending on the $V_{loc}$ value ($\eta_x$, $\eta_y$ and $\Gamma_x$, $\Gamma_y$ being fixed) saturation may occur within the $t$, $t^{1/4}$, $t^{1/2}$ or $\ln(t)$ regime.

The behavior of $G(t)$ for one vicinal surface can be determined from its average step density ($1/L$), together with its two energetic constants ($E_k$ and $A$) and hopping rates $\Gamma_x$, $\Gamma_y$. All parameters ($\eta_x$, $\eta_y$ and $\Gamma_x$, $\Gamma_y$) are temperature and a priori also $L$ dependent. The step density is given by the miscut $\varphi$ (angle between the terrace plane and the surface plane): $1/L = \sin(\varphi)/h$ with $h$ the elementary step height. We note $\theta$ the step orientation (angle between the step direction and the closest surface dense atomic row). Within the capillary wave approach, surface stiffnesses are related to energetic parameters ($E_k$ and $A$) according to [1]:

$$\eta_y = \frac{kT}{b^2(\theta, T)} + \frac{1}{L^2}\frac{d^2\delta(\theta, T)}{d(\tan(\theta))^2} \quad \text{and}$$
$$\eta_x = \frac{6\delta(\theta, T)}{L^4}, \quad (27)$$

where

$$\delta(\theta, T) = \frac{\pi^2}{6}kTb^2(\theta, T)\lambda^2 \quad \text{with}$$
$$\lambda = \frac{1}{2}\left[1 + \sqrt{1 + \frac{4A}{kTb^2(\theta, T)}}\right] \quad (28)$$

and where $b^2(\theta, T)$ is the step diffusivity whose expression for $\theta \neq 0$ in [15] reduces for $\theta = 0$ to:

$$b^2(\theta = 0, T) = \frac{1}{\cosh(E_k/kT) - 1}. \quad (29)$$

Note that when neglecting the small second term in $\eta_y$, $T_R$ is given by:

$$T_R = \frac{2}{\pi}\sqrt{\eta_x\eta_y}, \qquad \frac{A}{kT_R b^2(\theta, T_R)} = \frac{(L^2-1)^2 - 1}{4}. \quad (30)$$

Similarly, the energetic anisotropy can be approximated by:

$$\frac{\eta_y}{\eta_x} \approx \frac{kTL^4}{\pi^2 A b^2(\theta, T)}, \quad (31)$$

that is an increasing function of $L$ and a decreasing function of $T$. Table 2 sums up the $E_k$ and $A$ values for Cu(0 0 1) and Cu(1 1 1) vicinals that we used to evaluate $\eta_x$ and $\eta_y$.

Concerning hopping rates, their relation with diffusion parameters (nature and concentration of diffusing species, activation energies for diffusion) is less clear and is still a wide matter of debate [26]. Considering the $C_{ad}$ adatom concentration along step edges at thermal equilibrium, and their diffusion factor along step edges, the hopping rate for adatoms along straight step edges is:

$$\Gamma_y^{ad} = C_{step}^{ad} D_{step}^{ad} \quad \text{with}$$
$$D_{step}^{ad} \approx v_y \exp\left(-\frac{E_y^{Diff}}{kT}\right), \quad (32)$$

where $E_y^{Diff}$ is the activation energy for adatom diffusion along straight step edges. Detachment along the step edge of one atom from one kink

Table 2
Geometric and energetic parameters for vicinal surfaces of Cu

|  | Vicinals of Cu(0 0 1) | | | Vicinal of Cu(1 1 1) |
|---|---|---|---|---|
|  | (1 1 5) [34] | (1 1 11) [7] | (1 1 19) [2] | (21 21 23) [2] |
| $\varphi$ | 15.79° | 7.32° | 4.26° | 2.49° |
| Terrace width | 2.5 at.u. | 5.5 at.u. | 9.5 at.u. | 65/3 at.u. |
| $L$ | 0.66 nm | 1.41 nm | 2.42 nm | 4.78 nm |
| $E_k$ | 1430 ± 20 K | 1430 K | 1484 K | 1310 K |
| $A$ | 65 ± 5 K* | 70 K | 82 K | 71 K |

*In our Monte-Carlo analysis step–step interaction was restricted to first neighbors, giving a $A_{MC}$ value of 100 K. Eqs. (27) and (28) are valid for a model where interaction between every pair of steps is considered [1]. By means of the capillary wave model (interaction between all steps), one gets $A = 65$ K $\approx 6A_{MC}/\pi^2$.

giving formation of two kinks, the adatom concentration at thermal equilibrium is:

$$C_{\text{step}}^{\text{ad}} = \exp\left(-\frac{2E_k}{kT}\right) \quad (33)$$

and the total activation energy for $\Gamma_y^{\text{ad}}$ is equal to $2E_k + E_y^{\text{Diff}}$.

Displacement of kinks is responsible of fluctuations of step edge positions. Measurements of the time correlation of step edge positions are thus sensitive to kink diffusion [3]. Considering that kink diffusion occurs by detachment–attachment of atoms from one kink to another, the total activation energies for $\Gamma_y$ was found equal to $3E_k + E_y^{\text{Diff}}$ when adatom diffusion along a kinked step is considered [2,27,28], or to $2E_k + E_y^{\text{Diff}}$ by modeling the step edge as a 1-d solid-on-solid surface [26]. We could also consider that $\Gamma_y$ is the product of the concentration $C_k = b^2(T) \approx 2\exp(-E_k/kT)$ of the diffusing species (kinks at steps) by a kink diffusion factor in the $y$ direction with an activation energy $E_y^{\text{Diff}\,k}$.

$$\Gamma_y = C_k D_y^k \quad \text{with} \quad D_y^k \approx v_y \exp\left(-\frac{E_y^{\text{Diff}\,k}}{kT}\right). \quad (34)$$

A kink displacement of one site along $y$ occurs when an adatom detaches or attaches to the kink. The activation energy associated to $D_y^k$ is equal here to the attachment–detachment activation energy $E_y^{\text{Diff}\,k} = 2E_k + E_y^{\text{Diff}}$, giving a total activation energy for $\Gamma_y$: $3E_k + E_y^{\text{Diff}}$ in agreement with [2]. In the following, we keep this expression for numerical application. Other expressions for $\Gamma_y$ (according to [26] and/or including an additional barrier for adatom diffusion across kinks) could be considered and would alter only quantitatively our $(T–L)$ and $(T–t)$ diagrams. Fig. 6 gives a scheme of the evolution of the system energy during a kink diffusion event and sum up our notations.

In the $x$ direction, one expression for the parameter $\Gamma_x$ we have introduced can be obtained using the dimensional analysis as in the Case D (diffusion from step to step) of Pimpinelli et al. [4], giving:

$$\Gamma_x \approx \frac{C_{\text{ter}}^{\text{ad}} D_{\text{ter}}^{\text{ad}}}{L}, \quad (35)$$

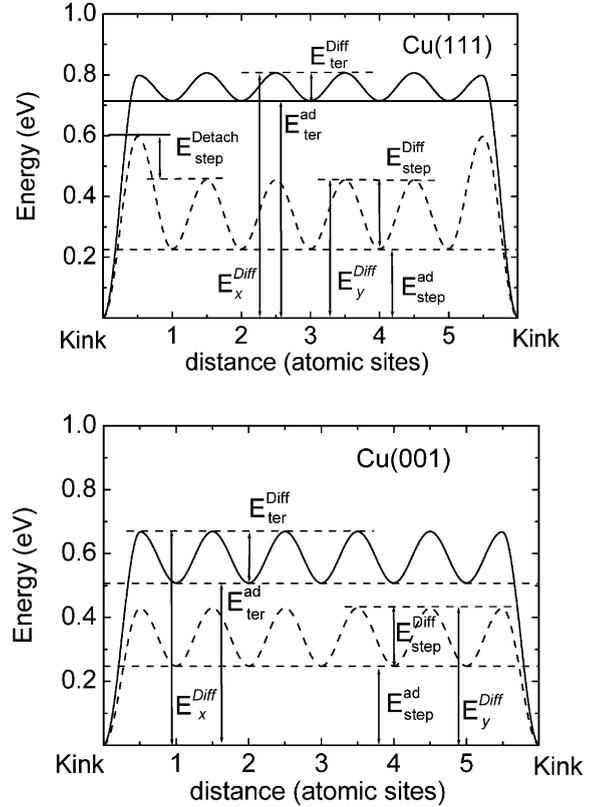

Fig. 6. Energy of the system during a kink diffusion event by atom detachment–attachment from kink sites. Kink diffusion is sensitive to the total barrier height. Dashed lines: diffusion along steps ($\Gamma_y$) (by adatom detachment–attachment from kinks from/onto the same step edge), solid lines: diffusion along the $x$ direction ($\Gamma_x$) (by adatom detachment–attachment from kinks onto/from terraces). Parameters are given in Table 3.

where $C_{\text{ter}}^{\text{ad}}$ is the adatom concentration over the terraces and $D_{\text{ter}}^{\text{ad}}$ their diffusion factor on the terraces. In this relation the factor $1/L$ was introduced as the probability for one detaching atom from one kink at a given step to attach to another kink at an adjacent step [4]. Introducing this expression into our Eq. (21), one gets the very same expression for $G(t)$ than Eq. (3.24) of [2]:

$$G(t) \approx \left[\frac{C_{\text{ad}} D_{\text{ter}}^{\text{ad}}}{L} \frac{kT}{\eta_y} t\right]^{1/2} \quad \text{with}$$

$$b^2(\theta, T) \approx \frac{kT}{\eta_y} \quad \text{and} \quad D_{\text{ter}}^{\text{ad}} = v_x \exp\left(-\frac{E_{\text{ter}}^{\text{Diff}}}{kT}\right)$$

(36)

with $E_{\text{ter}}^{\text{Diff}}$ the adatom diffusion energy on the terraces. $E_{\text{ter}}^{\text{ad}}$ being the energy to create an adatom on the terraces from a kink, their concentration is: $C_{\text{ter}}^{\text{ad}} = \exp(-E_{\text{ter}}^{\text{ad}}/kT)$ and the total activation energy for $\Gamma_x$ is in that case: $E_k + E_{\text{ter}}^{\text{ad}} + E_{\text{ter}}^{\text{Diff}}$.

A similar expression can be obtained by a way, consistent with our derivation of $\Gamma_y$. Writing the hopping rate as the product of concentration species by their diffusion factor in the $x$ direction, one gets:

$$\Gamma_x = \frac{C_k}{L} D_x^k \quad \text{with} \quad D_x^k \approx v_x \exp\left(-\frac{E_x^{\text{Diff } k}}{kT}\right) \quad (37)$$

where $C_k$ is the very same kink concentration as for $\Gamma_y$ and $D_x^k$ their diffusion factor in the $x$ direction. $C_k/L = b^2(T)/L \approx [2\exp(-E_k/kT)]/L$ is now the concentration of diffusing species within one unit cell ($a \times L$) for diffusion in the $x$ direction, the factor $1/L$ reflecting here the geometrical anisotropy. The activation energy for $D_x^k$ is the activation energy for detachment or attachment of adatoms onto/from the terraces. Without a Schwoebel barrier at steps, it is equal to $E_x^{\text{Diff } k} = E_{\text{ter}}^{\text{ad}} + E_{\text{ter}}^{\text{Diff}}$, giving a total activation energy for $\Gamma_x = E_k + E_{\text{ter}}^{\text{ad}} + E_{\text{ter}}^{\text{Diff}}$. As for $\Gamma_y$, other expressions of the activation energy for $\Gamma_x$ can be used that alter only quantitatively the following ($T$–$L$) and ($T$–$t$) diagrams.

For numerical application, we used results of effective medium theory (EMT) calculations by Stoltze [29] where a full set of parameters for vicinals of Cu on static substrates is available. Table 3 sums up these results. A common prefactor $v_x = v_y = 10^{13}$ is arbitrarily chosen. Using the above expressions for $\eta_x$, $\eta_y$ (27), $\Gamma_x$ (35), and $\Gamma_y$ (34), a $T$–$L$ diagram, giving the domains where the energetic or the diffusion anisotropy dominates the systems defined by the above parameters, can be obtained (see Fig. 7). This figure shows that the behavior of a vicinal surface is dominated by its energetic anisotropy at low $T$ and by its hopping anisotropy at high $T$. The limit between the two domains is plotted together with the roughening temperature $T_R(L)$ according to Eq. (30). For low $T$ values, where $\eta_y/\eta_x \ll \Gamma_y/\Gamma_x$ no extended $t^{1/2}$ regime is present. It is always the case in the flat phase ($T < T_R$).

Fig. 8 shows for $L = 65/3$ at.u. (as for Cu(21 21 23)), a wide extension of the $t^{1/4}$ regime and in contrast a much narrower extension of the $t^{1/2}$ regime, vanishing for low $T$, before logarithmic saturation occurs. Note that the plotted limits correspond to smooth transitions between regions where well-defined $t^{1/n}$ regimes are found. The ($t$–$L$) diagram of Fig. 9 (where values of $E_x^{\text{Diff } k}$ and $E_y^{\text{Diff } k}$ resulting from the fit of data on vicinals of

Table 3
EMT adatom creation energies [29] and activation energies for kink diffusion along step edges (i.e. adatom detachment from kinks to step edge) ($E_y^{\text{Diff}}$) and from one step to another step (i.e. adatom detachment from kinks to terraces) ($E_x^{\text{Diff}}$) as obtained from analysis of experimental data of [2,8,30]. All energies are in eV

|  |  | Cu(1 1 1) | Cu(0 0 1) |
|---|---|---|---|
| Creation of adatom along step edges $\approx 2E_k$ | $E_{\text{step}}^{\text{ad}}$ | 0.23 [2] | 0.247 [1,2] |
| Activation energy for diffusion along step edges (EMT) | $E_{\text{step}}^{\text{Diff}}$ | 0.226 [29] | 0.247 [29] |
| Prefactor | $v_x = v_y$ | $5 \times 10^{12}$ | $1.5 \times 10^{13}$ |
| Activation energy for kink diffusion along step edges (this work from data of [2,8,30]) | $E_y^{\text{Diff } k}$ | 0.60 | 0.43 |
| Additional barrier at kink site for detachment along step edges: $E_y^{\text{Diff } k} - (E_{\text{step}}^{\text{ad}} + E_{\text{step}}^{\text{Diff}})$ | $\delta E_{\text{step}}$ | 0.144 | $-0.064$ |
| Creation of adatom onto terraces (EMT) | $E_{\text{ter}}^{\text{ad}}$ | 0.714 [29] | 0.507 [29] |
| Activation energy for diffusion onto terraces (EMT) | $E_{\text{ter}}^{\text{Diff}}$ | 0.094 [29] | 0.425 [29] or 0.21 [32] |
| Activation energy for kink diffusion on terraces (this work from data of [2,8,30]) | $E_x^{\text{Diff } k}$ | 0.80 | 0.67 |
| Additional barrier at kink site for detachment along step edges: $E_x^{\text{Diff } k} - (E_{\text{ter}}^{\text{ad}} + E_{\text{ter}}^{\text{Diff}})$ | $\delta E_{\text{ter}}$ | $-0.008$ | $-0.262$ or $-0.047$ |

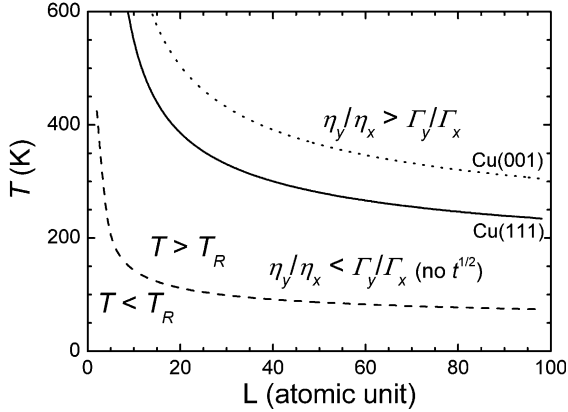

Fig. 7. $T$–$L$ diagram showing the domains where the energetic or the diffusion anisotropy dominates the system. (—). $v_y = v_x = 10^{13}$ s$^{-1}$. For vicinals of Cu(1 1 1), parameters are $E_k = 1310$ K, $A = 71$ K, total activation energy for $\Gamma_y$ ($3E_k + E_y^{\text{Diff}}$) = $E_k + 0.456$ eV and for $\Gamma_x$ ($E_k + E_{\text{ter}}^{\text{ad}} + E_{\text{ter}}^{\text{Diff}}$) = $E_k + 0.808$ eV according to Stoltze [29]. Dotted line: vicinals of Cu(0 0 1), $E_k = 1430$ K, $A = 70$ K, total activation energy for $\Gamma_y = E_k + 0.494$ eV and for $\Gamma_x = E_k + 0.932$ eV. Also given (dashed line) the roughening temperature $T_R(L)$ according to Eq. (30). The relatively small differences between the energetic parameters of vicinals of Cu(0 0 1) and Cu(1 1 1) (see Table 2) makes the $T_R$ curves similar for both vicinals.

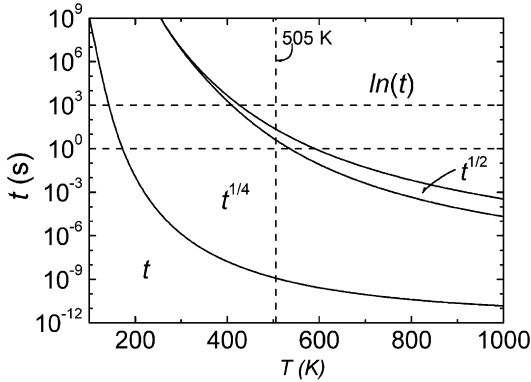

Fig. 8. $t$–$T$ diagram showing the extension of the various regimes of $G(t)$ for $L = 65/3$ at.u. ($L = 4.78$ nm, for a vicinal of Cu(1 1 1)). Other parameters are: $E_k = 1310$ K, $A = 71$ K, $v_y = v_x = 10^{13}$ s$^{-1}$, activation energies for kink diffusion: $E_y^{\text{Diff }k} = 0.456$ eV and $E_x^{\text{Diff }k} = 0.808$ eV according to Stoltze [29]. A 1-d Langevin approach is valid (Eq. (19)) for times within the $t^{1/4}$ domain. Horizontal dashed lines gives one typical measurement window ($1 < t < 1000$ s).

Cu(1 1 1) are used, see Section 5) shows for $T = 505$ K that the intermediate $t^{1/2}$ regime van-ishes for low $L$ values. Disappearing of the $t^{1/2}$ regime corresponds to the transition from a dominant stiffness anisotropy $\eta_y/\eta_x \gg \Gamma_y/\Gamma_x$ for wide terraces or high $T$ to a dominant hopping anisotropy $\eta_y/\eta_x \ll \Gamma_y/\Gamma_x$ for narrow terraces or low $T$. A 1-d Langevin approach can be used (Eq. (19)) within the $t^{1/4}$ domain.

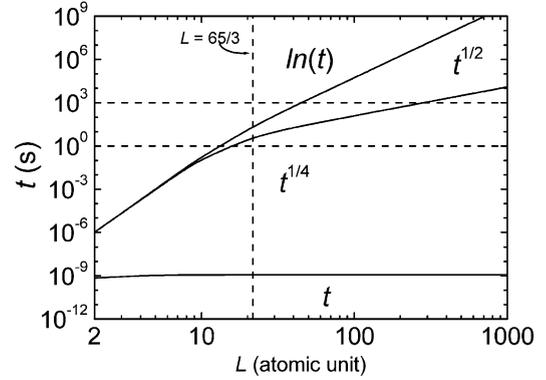

Fig. 9. $t$–$L$ diagram showing the extension of the various regimes of $G(t)$ for $T = 505$ K. Other parameters are: $E_k = 1310$ K, $A = 71$ K, $v_y = v_x = 5 \times 10^{12}$ s$^{-1}$, $E_y^{\text{Diff }k} = 0.6$ eV and $E_x^{\text{Diff }k} = 0.8$ eV. A 1-d Langevin approach is valid (Eq. (19)) for times within the $t^{1/4}$ domain. Horizontal dashed lines gives one typical measurement window ($1 < t < 1000$ s).

## 5. Fit of data for vicinals of Cu(0 0 1) and Cu(1 1 1)

Within our 2-d Langevin formalism, one could tentatively fit available data for vicinals of Cu(0 0 1) [2,8] and Cu(1 1 1) [30] using the original $G(t)$ curves. $E_k$ and $A$ values of Table 2 are used. Concerning Cu(1 1 1), measurements of $G(t)$ were performed on the Cu(21 21 23) surface between 305 and 500 K [30]. $G(t)$ was found to exhibit an apparent power law with $n = 1/4$ exponent. However, fitting these data, in the low temperature range (305–440 K) is not enough to fix accurately $D_y^k$ and $D_x^k$. The difference in activation energy $E_x^{\text{Diff }k} - E_y^{\text{Diff }k}$ can be much more precisely fixed considering the transition between the $t^{1/4}$ and the $t^{1/2}$ regime observed around $t = 30$ s at $T = 505$ K for 60–150 nm distant steps on Cu(1 1 1) [31]. In our fitting procedure, an intermediate value of 100 nm for the terrace width is taken. To reproduce a transition at $t = 30$ s, a value of 0.2 eV for $E_x^{\text{Diff }k} - E_y^{\text{Diff }k}$ is obtained that is found weakly dependant on $D_y^k$. Thus, fitting the set of data for $G(t = 1 \text{ s})$ with our Eq. (11) (see Figs. 10 and 11) and assuming the very same frequency prefactor for $D_x^k$ and $D_y^k$, one gets: $v_x = v_y = 5 \times 10^{12}$, $E_y^{\text{Diff }k} = 0.60$ eV and $E_x^{\text{Diff }k} = E_y^{\text{Diff }k} + 0.2 = 0.80$ eV. Concerning vicinals of Cu(0 0 1), $G(t)$ was measured for Cu(1 1 19) (310–360 K) and Cu(1 1 79) (390–600 K) [2,8]. Fitting $G(t)$ for $t = 0.03$, 1 and 4 s with our Eq. (11) (see Fig. 10) allows getting reasonable absolute values of $G(t)$ and of the apparent time exponents with the parameters: $v_x = v_y = 1.5 \times 10^{13}$ and $E_y^{\text{Diff }k} = 0.43$ eV and $E_x^{\text{Diff }k} = 0.67$ eV. The quality of the fit of the whole data indicates that these values can be considered as reasonable. Calculation of valid error bars needs a fit including the full set of original data.

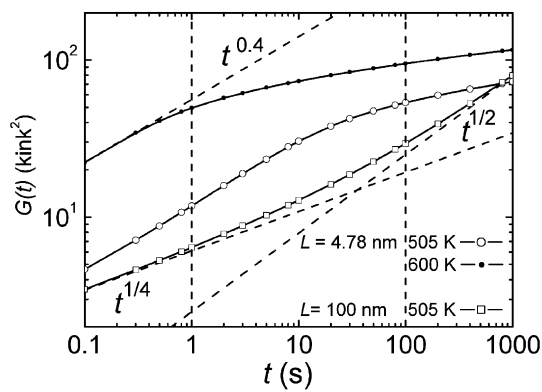

Fig. 11. $G(t)$ for $L = 65/3$ at.u. $= 4.78$ nm, (as for Cu(21 21 23), $L = 4.78$ nm), other parameters are: $E_k = 1310$ K, $A = 71$ K, $v_y = v_x = 5 \times 10^{12}$ s$^{-1}$, $E_y^{\text{Diff }k} = 0.60$ eV and $E_x^{\text{Diff }k} - E_y^{\text{Diff }k} = 0.2$ eV. Curves: (—O—O—) $T = 505$ K and (—●—●—) 600 K. Logarithmic divergence is observed after the $t^{1/4}$ regime (without an intermediate extended $t^{1/2}$ regime). $G(t)$ for widely separated steps: $L = 100$ nm $= 453$ at.u. (—□—□—) $T = 505$ K. A $t^{1/4}$–$t^{1/2}$ transition is obtained. Vertical dashed lines indicate the measurement window in [2].

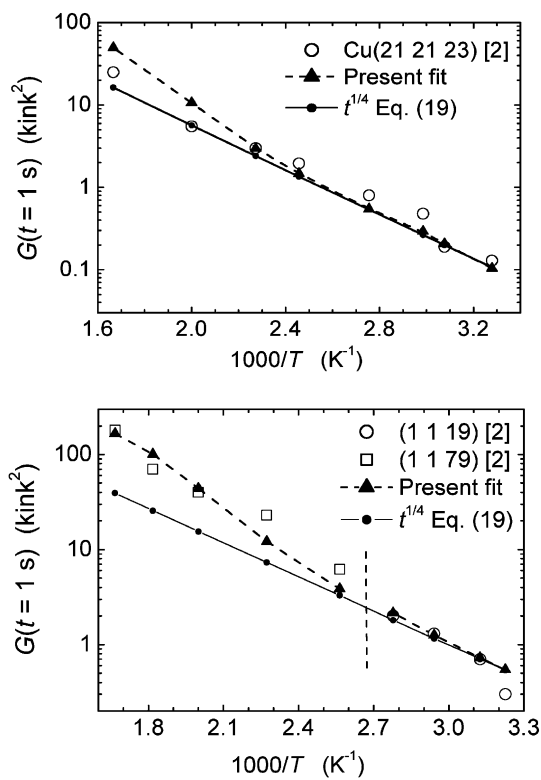

Fig. 10. Top: Cu(21 21 23), (□) $G(t = 1 \text{ s})$ versus temperature [2,30]. Bottom: vicinals of Cu(0 0 1). (O, □) experimental data from [8]. (- -▲- -) fit of the set of experimental data using Eq. (11). For Cu(21 21 23) parameters are: $L = 65/3$ at.u. $= 4.78$ nm, $E_k = 1310$ K, $A = 71$ K and one gets $v_y = v_x = 5 \times 10^{12}$ s$^{-1}$, $E_y^{\text{Diff }k} = 0.60$ eV and $E_x^{\text{Diff }k} = E_y^{\text{Diff }k} + 0.2 = 0.80$ eV. For vicinals of Cu(0 0 1), parameters are: $E_k = 1430$ K, $A = 70$ K and one gets $v_y = v_x = 1.5 \times 10^{13}$ s$^{-1}$, $E_y^{\text{Diff }k} = 0.43$ eV and $E_x^{\text{Diff }k} = 0.67$ eV. The vertical dashed line separates (1 1 19) and (1 1 79) data. (—●—●—) $t^{1/4}$ power laws according to Eq. (19) with the same parameters.

Also given in Fig. 10 values of $G(t = 1 \text{ s})$ obtained in the 1-d Langevin approach ($\Gamma_x = 0$, Eq. (19)). Differences with the result of the 2-d Langevin calculation shows that for the considered values of the parameters, a contribution of adatom attachment–detachment from steps is present at high temperature.

One could add further comments on the activation energies that we have obtained. The activation energy for $D_x^k$ must be at least higher than the sum of the creation energy of an adatom on the terrace from one kink ($E_{\text{ter}}^{\text{ad}}$) and its activation energy for diffusion on terraces ($E_{\text{ter}}^{\text{Diff}}$) as already considered in [31]. Along the steps, $E_y^{\text{Diff }k}$ must be at least higher than the sum of the creation energy for an adatom along the step edge from a kink ($E_{\text{step}}^{\text{ad}}$) and its activation energy for diffusion along the steps ($E_{\text{step}}^{\text{Diff}}$) (see Fig. 6). An additional energy barrier $\delta E_{\text{step}} = E_y^{\text{Diff }k} - (E_{\text{step}}^{\text{ad}} + E_{\text{step}}^{\text{Diff}})$ and $\delta E_{\text{ter}} = E_x^{\text{Diff }k} - (E_{\text{ter}}^{\text{ad}} + E_{\text{ter}}^{\text{Diff}})$ due to detachment of adatoms from kinks towards step edges or terraces or to a Schwoebel barrier (diffusing adatoms in the $x$ direction must cross steps) may be also considered. Although a simple diffusion path is considered here, more complex diffusion path including some asymmetry between detachment–attachment at kinks, adatom-kink crossing along step edges as well as for adatoms ascending or descending one step could be introduced.

At first, it can be noticed that $v_x$ and $v_y$ are typical pre-exponential factors for diffusion lying in the $10^{12}$–$10^{13}$ range. The $E_y^{\text{Diff }k}$ (0.60 eV) and $E_x^{\text{Diff }k}$ (0.20 eV) values we obtain for vicinals of Cu(1 1 1) can be compared with the results of EMT calculations of Stoltze [32]. Table 3 sum up the obtained activation energies for kink diffusion and the EMT results for its components. The obtained values compare well with the theoretical ones, giving an extra energy $\delta E_{\text{step}}$ of only 0.14 eV that could be attributed to the kink detachment and almost 0 for $\delta E_{\text{ter}}$.

For vicinals of Cu(0 0 1), comparison with Stoltze's values gives an almost negligible value of $\delta E_{\text{step}}$ and a surprisingly negative value of −0.26 eV for $\delta E_{\text{ter}}$. We must be cautious about this disagreement, as the available data do not allow a very accurate fit of $E_x^{\text{Diff }k}$ (for Cu(1 1 1) a much better sensitivity on this parameter is obtained considering the observed $t^{1/4}$–$t^{1/2}$ transition). Note however that for diffusion by an exchange mechanism on Cu(0 0 1) on a dynamic substrate, values of the free energy $E_y^{\text{Diff }k}$ as low as 0.2 eV has been obtained by a coupled EMT Monte-Carlo approach [32], for which for $\delta E_{\text{ter}}$ would be almost 0. However, this last value of $E_y^{\text{Diff }k}$ appears controversial as ab initio calculation gives an activation energy of 0.96 eV for the exchange mechanism [33]. Comparison with other theoretical results can be also done through the extensive review of Giesen in [2]. Beyond these values, fits including Schwoebel barriers and thus different expression for $\Gamma_x$ and $\Gamma_y$ have to be done. Our fit shows that reasonable order of magnitudes for adatom formation and their diffusion (along steps and onto terraces) are obtained using our 2-d Langevin analysis.

Finally, some more comments can be done on the $t^{1/4}$–$t^{1/2}$ transition observed around 500 K [2,31] on one vicinal of Cu(1 1 1) with 100 nm wide terraces. This behavior can be compared to the one of the nominal Cu(21 21 23) surface, where a $t^{1/4}$ regime was observed below $T = 500$ K and $t^{0.45}$ at 600 K for short times (0.1–10 s) [30]. For $L = 65/3$ (as for Cu(21 21 23)) at $T = 505$ K and using our parameters for Cu(1 1 1) vicinals, one gets the $G(t)$ curve given in Fig. 11. For this temperature, the domain where the $t^{1/2}$ regime would be present is so narrow that the logarithmic divergence is better seen after the initial $t^{1/4}$ regime. One further obtains for $T = 600$ K, an apparent exponent $1/n = 0.4$ (within the interval 0.1–3 s) close to the experimental value [30] and a logarithmic variation for $t$ above a few seconds. This indicates that only the temperature variation of the exponent at short times and not a $t^{1/4}$–$t^{1/2}$ transition *at one given temperature* can be pointed out on the nominal surface due to logarithmic divergence for long times. Conversely, the time domain for the $t^{1/2}$ behavior being much more extended for larger step–step distance (see Fig. 9), one finds, in agreement with the experiment, that the transition must be clearly observable for $L = 100$ nm.

## 6. Cu(1 1 5) STM study

As previously indicated, the 2-d Langevin approach is mandatory to analyze experiments for times higher than $kT/4\eta_x\Gamma_x$ (i.e. at high $T$ (see Figs. 7 and 8) or vicinals with narrow terraces where lower energetic or diffusion anisotropy can be expected). It is also the case for surfaces below the roughening transition. The present 2-d Langevin approach is thus the appropriate tool for analyzing our low temperature measurements of $G(t)$ on the Cu(1 1 5) vicinal surface, which energetic parameters has been recently obtained from the analysis of spatial correlation function measurements [34].

### 6.1. Experimental

The Cu(1 1 5) sample has been studied by variable temperature STM (VT-STM). Our VT-STM is a commercial Omicron. Electrochemically etched tungsten tips were used. A resistive PBN plate on the back side of the sample and inside the sample holder allowed sample heating. Depending on tip stability, surface quality and temperature, tip-scanning speed between 40 and 230 nm/s were used and data taking time for $20 \times 20$ nm$^2$ STM pictures varied from 25 s to 2.5 min.

The sample has been spark cut from a single crystal ($9 \times 3$ mm$^2$, 2 mm thick) and mechanically polished. The surface orientation (miscut angle 15.79°) is along the (1 1 5) direction within 0.15° as shown by X-ray Laue diffraction. Under UHV, surfaces are cleaned by cycles of Ar$^+$ bombardment (typically: 3 µA, 600 eV, 1 h) followed by short annealing above 850 K. The lattice constant is $a_0 = 0.360$ nm. Steps are oriented along the $[1\bar{1}0]$ dense direction, atoms at step edges are separated by the next nearest neighbor distance $a = a_0/\sqrt{2} = 0.255$ nm. The average terrace width (2.5 atoms wide) is $L_0 = 2.5, a = 0.636$ nm. The step–step average distance and the step height are respectively $L = 0.66$ nm and $h = 0.18$ nm (see Fig. 1).

### 6.2. Experimental results and analysis of STM images

On the Cu(1 1 5) surface, steps are very close and a low apparent surface corrugation results. A high resolution (0.07 nm/pixel, i.e. 3–4 pixels per atomic distance) is necessary to reach the atomic resolution allowing observation of kinks at steps without digitalization noise. It is worth noting that the surface corrugation is much lower for this closely spaced step surface than for vicinals with higher index, like Cu(1 1 11) [15].

In a recent contribution devoted to a comparative study of the thermal roughness of the Cu(1 1 5) and the Cu–Pd(17%)(1 1 5) alloy surface, VT-STM images were analyzed [34]. The second moment of the terrace width distribution was found to be lower than the universal value $4/\pi^2$ [1]. It is a first indication that the Cu(1 1 5) surface at room temperature is below its roughening transition. The observed saturation of $G(m, 0)$ is also an indication that the surface is below $T_R$. Measurements of the correlation function of step edge displacements $G(m, 0)$, in the fast scan direction where the time dependence can be neglected, allowed characterizing the surface roughness at three temperatures: 300, 325 and 365 K.

The temperature range ($T < T_R$) and the time dependence in $G(0, y, t)$ complicate the data analysis. As previously reported in [34], considering only $G(m, 0)$ is not enough to determine unambiguously $\eta_x$, $\eta_y$ and $V_{loc}$ for $T < T_R$. We proceed as follows. Monte-Carlo simulation (see [34]) together with the capillary wave approach, where surface stiffnesses for $T > T_R$ can be obtained from (27) (see also the solution of the lattice Calogero Sutherland model [14]) are used. Parameters are obtained by fitting together the $G(m, 0)$ data for the three temperatures. A Monte-Carlo fit, where only elastic step–step interactions restricted to adjacent steps are considered, gives $E_k = 1430$ K and $A_{MC} = 100$ K. Within the capillary wave approach, where elastic interactions between all pairs of steps is considered, $E_k = 1430$ K and $A_{Cap} = 65$ K values allow reproducing Monte-Carlo simulations for $T > T_R$. Fixing the $E_k$ (1430 K) and $A_{MC}$ (100 K) values, $G(0, y)$ can be further obtained by MC simulations for $T < T_R$. We have now enough data to get, by means of Eq. (12) (see also [1]), values of $\eta_x$, $\eta_y$ and $V_{loc}$ reproducing the MC results ($G(m, 0)$ and $G(0, y)$ for each temperature). So-obtained parameters are reported in Fig. 12 and in Table 4 for the three temperatures. As noticed in [1] it is noteworthy that the $\eta_x$ expression remains valid below $T_R$ whereas $\eta_y$ is strongly affected.

In summary, measurements of the step correlation function ($G(m, 0)$) for three temperatures below $T_R$ allowed to get the energetic parameters of the vicinal surface: kink energy: $E_k = 1430$ K and step–step interaction constant: $A = 65$ K. A roughening temperature ($T_R$) of 380 K is further deduced [34]. Calculation of $G(0, y, 0)$ with the obtained parameters gives much lower values than the measured quantity $G(0, y, t)$. This confirms that the time dependence (each data being separated by the time necessary to scan one picture line) cannot be neglected in these measurements and that images are indeed space–time mixed.

In order to separate spatial and temporal dependences, time images were recorded by repetitively scanning one single line along $x$ (see Fig. 13).

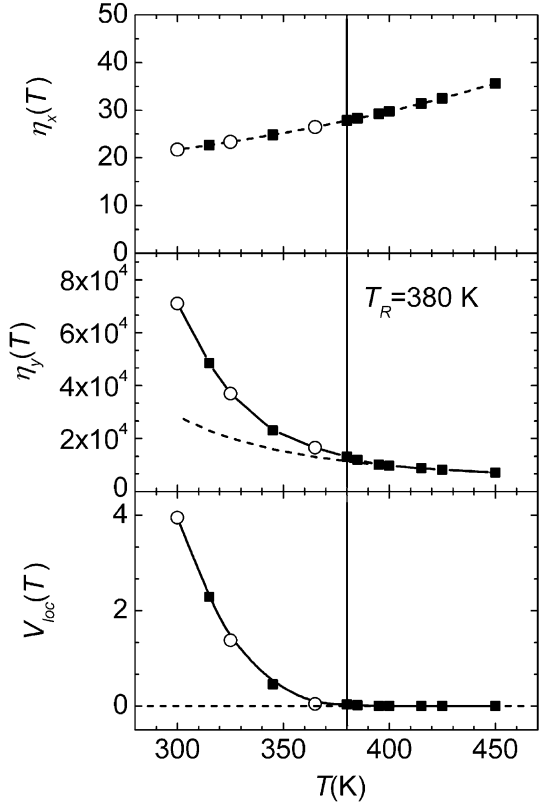

Fig. 12. (■) Temperature variations of $\eta_x$, $\eta_y$ and $V_{loc}$ (in Kelvin) fitting $G(m,0)$ and $G(0,y)$ obtained by MC simulations. (○) for measurement temperatures ($T < T_R$). Parameters of the MC simulation (elastic step interaction restricted to adjacent steps) are: $E_k = 1430$ K and $A_{MC} = 100$ K (system size: 76 steps $\times$ 1300 $a$). For $T > T_R$ the capillary approach (Eq. (9), elastic interaction between every pair of steps) gives $A_{Cap} = 65$ K. Continuous line: guide to the eye, dashed lines: $\eta_x$ and $\eta_y$ according to Eq. (27) (see also [1]).

For $T = 300$ K, very few step displacements are observed whereas for $T = 365$ K rapid fluctuations are visible. From statistical analysis on several images (see Table 5), the time correlation function $G(0,0,t)$ is obtained. Results are given in Fig. 14 for the three temperatures. $G(0,y,t)$ and $G(t)$ appear very similar (see Fig. 15). This experimentally confirms that the time dependence cannot be neglected in $G(0,y,t)$. For short times, fitting $G(t)$ with a power law $t^{1/n}$ gives an apparent exponent with $n = 8$, much lower than the exponent 1/4 measured for isolated steps in the diffusion along step edge regime (room temperature for Cu). Moreover at larger times $G(t)$ tends to saturate like $G(m,0)$ or $G(0,y,t)$. The following analysis shows that step–step interactions are indeed responsible for so low time fluctuations.

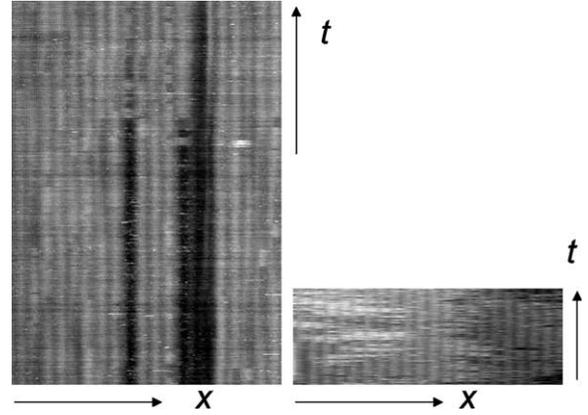

Fig. 13. VT-STM time images of Cu(1 1 5). Left: $T = 300$ K (14 nm $\times$ 80 s, 152 $\times$ 400 pixels$^2$), scan speed along the $x$ direction: 140 nm/s. Right: $T = 365$ K (14 nm $\times$ 20 s, 152 $\times$ 200 pixels$^2$), scan speed 280 nm/s.

The energetic parameters being known, time and spatio-temporal correlation functions can be fitted. Our measurements being in the low $T$ range ($T < 400$ K), $\Gamma_x$ can be neglected (see Fig. 10). $\Gamma_y$ values allowing a good fit of $G(t)$ (Fig. 14) and $G(0,y,t)$ (Fig. 15) result and are reported in

Table 4
Parameters fitting the spatial (Eq. (12)) and spatio-temporal (Eq. (9)) correlation functions

|  | 300 K | 325 K | 365 K |
|---|---|---|---|
| $\eta_x$ (K) | 21.8 | 23.4 | 26.5 |
| $\eta_y$ (K) | $7.1 \times 10^4$ | $3.7 \times 10^4$ | $1.65 \times 10^4$ |
| $V_{loc}$ (K) | 3.95 | 1.38 | 0.04 |
| $\Gamma_x$ (s$^{-1}$) | 0 | 0 | 0 |
| $\Gamma_y$ (s$^{-1}$) | $(1.0^{+0.7}_{-0.4}) \times 10^2$ | $(3.0^{+2}_{-1}) \times 10^3$ | $(6^{+4}_{-2}) \times 10^4$ |

Values of $\eta_x$, $\eta_y$ and $V_{loc}$ are obtained so as to reproduce MC simulations for $E_k = 1430$ K and $A_{MC} = 100$ K.

Table 5
Number of analyzed data for $G(t)$ measurements

|  | 300 K | 325 K | 365 K |
| --- | --- | --- | --- |
| Number of images | 8 | 4 | 4 |
| Size of images | 14 nm × 100 s | 9 nm × 100 s | 18 nm × 100 s |
| Data/image | 152 × 400 pixels$^2$ | 52 × 400 pixels$^2$ | 109 × 750 pixels$^2$ |

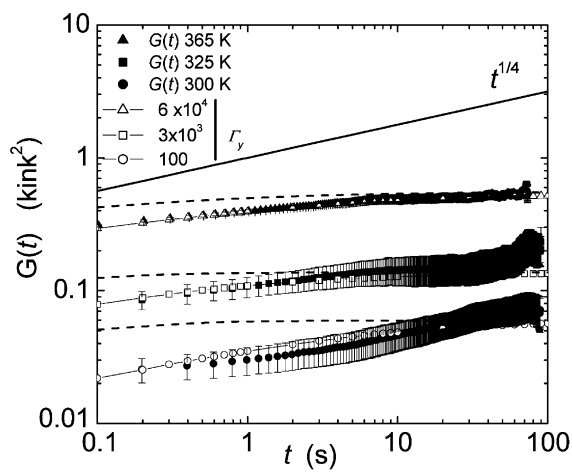

Fig. 14. Measured time correlation function, (●) $T = 300$ K, (■) 325 K, and (▲) 365 K. (○, □, △): Fit of $G(t)$ for the three temperatures (see Section 5 and parameters of Table 4). Dashed lines give $G(t)$ according to Eq. (11) with parameters ($E_k = 1430$ K, $A = 65$ K, $v_y = v_x = 1.5 \times 10^{13}$ s$^{-1}$, $E_y^{\text{Diff } k} = 0.43$ eV and $E_x^{\text{Diff } k} = 0.67$ eV) adjusted on Cu(1 1 19) and Cu(1 1 79) data [2,8]. A $t^{1/4}$ power law (upper line) is shown for comparison.

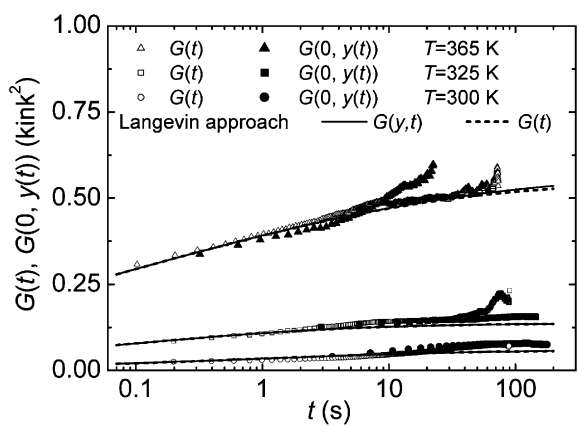

Fig. 15. Compared measured space–time correlation $G(0, y, t)$ (black symbols) and time $G(0, 0, t) = G(t)$ correlation functions (open symbols). Fit of $G(t)$ (solid line) and $G(0, y, t)$ (dashed line) are also given (see Section 5). For the experimentally used tip-scanning speeds, $G(t)$ and $G(0, y, t)$ cannot be distinguished.

Table 4. Also shown in Fig. 14, values of $G(t)$ according to Eq. (11) with the values of the parameters we obtained for Cu(0 0 1) vicinals ($v_x = v_y = 1.5 \times 10^{13}$, $E_y^{\text{Diff } k} = 0.43$ eV and $E_x^{\text{Diff } k} = 0.67$ eV) that are clearly above our measurements.

Temperature variations of $D_x^k$ and $D_y^k$, as obtained by fitting (for each temperature of measurement) data on Cu(1 1 19) and Cu(1 1 79) [8], are reported together with present results on Cu(1 1 5) in Fig. 16. A better accuracy is obtained for $D_y^k$ at low $T$ and for $D_x^k$ at high $T$. $D_y^k$ for Cu(1 1 5) appears slightly lower than the values extracted from Cu(1 1 19). Note that a deviation of $D_y^k$ from a straight line for Cu(1 1 19) values at low $T$ is also visible. Cu(1 1 5) being a surface with very narrow terraces (strong step–step interaction), the kink diffusion factor could be lower than those of other vicinals with wider terraces. One has also to consider that the kink diffusion can be altered below the roughening transition (non-zero localization potential).

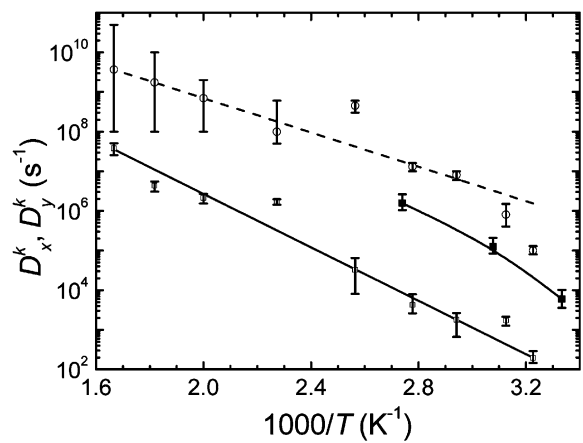

Fig. 16. Arrhenius plot of (○) $D_y^k$ and (□) $D_x^k$ for vicinals of Cu(0 0 1). Values are obtained by fitting with Eq. (11) $G(t)$ data for Cu(1 1 19) and Cu(1 1 79) from [8] for each measurement temperature. Error bars indicates the range for $D_x^k$ (resp. $D_y^k$), keeping the other diffusion parameter $D_y^k$ (resp. $D_x^k$) constant and allowing an acceptable fit of $G(t)$. Full line: $D_x^k$ and dashed line: $D_y^k$, for the parameters obtained for Cu(0 0 1) vicinals: $v_y = v_x = 1.5 \times 10^{13}$ s$^{-1}$, $E_y^{\text{Diff } k} = 0.43$ eV and $E_x^{\text{Diff } k} = 0.67$ eV. (■) $D_y^k$ values for Cu(1 1 5) assuming $\Gamma_x = 0$.

## 7. Conclusion

In summary, we have developed a 2-d Langevin formalism that provides handy relations to analyze fluctuations of a surface when matter conservation within the surface is considered. Energetic as well as diffusion anisotropies are considered. This allows analyzing space–time correlation functions of step displacements of vicinal surfaces that can be related to atomic energetic parameters ($E_k$ and $A$) and hopping rates. Concerning time correlation functions, approximate power law regimes are obtained for short times as observed for well-isolated steps or for weakly interacting steps at short times. Our model allows describing accurately the intermediate regimes. Temperature and terrace width domains where power law approximations apply are given. A logarithmic divergence for long times is ultimately reached in the rough phase ($T > T_R$) while saturation of $G(t)$ in the flat phase ($T < T_R$) is right obtained.

Within the model, two parameters for hoping rates of kinks (along and from one step to an adjacent one) are introduced. Following previous analysis [2,4], these parameters can be related to the kink concentration and the diffusion constants for kinks, which in turn depends on the activation energy for adatom formation and diffusion at step edges and on the terraces. An overall good agreement with theoretical values is obtained by fitting experimental data with the present 2-d Langevin model (except notably for diffusion across terraces for Cu(0 0 1) vicinals).

In addition to spatial correlation functions of step displacements previously analysed, we have measured the time correlation function. $G(t)$ on the flat ($T < T_R$) Cu(1 1 5) vicinal surface. $G(t)$, shows saturation for long times like spatial correlation functions. The analysis by the 2-d Langevin approach of the measured $G(t)$ and $G(0, y, t)$ correlation functions allowed to extract the kink diffusion factor for diffusion along steps ($D_y^k$) which values appear slightly lower than for vicinals of Cu(0 0 1) with wider terraces.

## Appendix A

### A.1. General formalism

We consider the following Langevin equation describing interacting step fluctuations:

$$\frac{\partial h_{m,y}}{\partial t} = \frac{1}{kT}\left(\Gamma_x \frac{\partial^2}{\partial m^2} + \Gamma_y \frac{\partial^2}{\partial y^2}\right)\left(\frac{\delta H}{\delta h_{m,y}}\right) + \xi(m,y,t). \quad (A.1)$$

Within a capillary wave model of a flat vicinal surface, the discrete Hamiltonian $H$ of an array of steps is [25]:

$$H = \sum_{m,y}\left(\frac{\eta_x}{2}(h_{m+1,y} - h_{m,y})^2 + \frac{\eta_y}{2}(h_{m,y+1} - h_{m,y})^2 + V(h_{m,y})\right). \quad (A.2)$$

The localization potential can be approximated by:

$$V(h_{m,y}) = 2V_{\text{loc}}(1 - \cos(2\pi h_{m,y})) \approx 4\pi^2 V_{\text{loc}} h_{m,y}^2. \quad (A.3)$$

For a local $(m_0, y_0)$ small variation of the surface height: $\delta h_{m,y} = \delta h_{m_0,y_0}\delta(m, m_0)\delta(y, y_0)$, the variation of the surface energy is:

$$\frac{\delta H}{\delta(h_{m,y})} = -\sum_{m,y}\left[\eta_x(h_{m+1,y} + h_{m-1,y} - 2h_{m,y}) + \eta_y(h_{m,y+1} + h_{m,y-1} - 2h_{m,y}) - \frac{\partial V(h_{m,y})}{\partial(h_{m,y})}\right]. \quad (A.4)$$

Introducing Eq. (A.4) in (6), and linearizing the differential equation by Fourier expansion,

$$h_{m,y}(t) = \frac{1}{\sqrt{N_x N_y}}\sum_{m,y} h_q(t)e^{i(q_x m + q_y y)},$$
$$\xi(m,y,t) = \frac{1}{\sqrt{N_x N_y}}\sum_{m,y}\xi_q(t)e^{i(q_x m + q_y y)}. \quad (A.5)$$

$N_x N_y$ being the system size, one gets:

$$\frac{\partial h_q(t)}{\partial t} = -\frac{1}{kT} a_q h_q(t) + \xi_q(t), \quad (A.6)$$

where $a_q$ is defined by:

$$a_q = \frac{4b_q}{kT}[\Gamma_x(1-\cos(q_x)) + \Gamma_y(1-\cos(q_y))]$$

with

$$b_q = \eta_x(1-\cos(q_x)) + \eta_y(1-\cos(q_y)) + 4\pi^2 V_{\text{loc}}. \quad (A.7)$$

Solution of the above differential equation is:

$$h_q(t) = e^{-a_q t} \int_0^t e^{a_q t'} \xi_q(t') \, dt' + h_q(t=0). \quad (A.8)$$

The second moment of the capillary mode of step edge fluctuations is further deduced:

$$\langle h_{-q}(t')h_q(t)\rangle = e^{-a_q(t'+t)} \int_0^{t'} \int_0^t e^{-a_q(t_1+t_2)}$$
$$\times \langle \xi_{-q}(t_1)\xi_q(t_2)\rangle \, dt_1 \, dt_2 + \cdots$$
$$+ \langle h_q(t')h_q(t=0)\rangle. \quad (A.9)$$

The correlation function for the thermal noise can be written as:

$$\langle \xi(m',y',t')\xi(m,y,t)\rangle$$
$$= -2\left(\Gamma_x \frac{\partial^2}{\partial m^2}\delta(m-m')\right.$$
$$\left. + \Gamma_y \frac{\partial^2}{\partial y^2}\delta(y-y')\right)\delta(t-t'). \quad (A.10)$$

The very same $\Gamma_y$ and $\Gamma_y$ constants as in (A.1) appear in the above equation in order to satisfy the energy equipartition for fluctuations at thermal equilibrium:

$$\langle H(h_q(t))^2\rangle = \frac{kT}{2}. \quad (A.11)$$

In the Fourier space, one obtains:

$$\langle h_{-q}(t)h_q(t)\rangle = \langle h_{-q}(t'=0)h_q(t=0)\rangle = \frac{kT}{2b_q},$$
$$\langle \xi_q(t')\xi_q(t)\rangle = 4[\Gamma_x(1-\cos(q_x))$$
$$+ \Gamma_y(1-\cos(q_y))]\delta(t-t'). \quad (A.12)$$

Introducing (A.12) into (A.9), one gets by integration:

$$\langle h_{-q}(t')h_{qy}(t)\rangle = \frac{e^{-a_q(t'-t)} - e^{-a_q(t'+t)}}{2a_q}$$
$$\times 4[\Gamma_x(1-\cos(q_x))$$
$$+ \Gamma_y(1-\cos(q_y))] + \cdots$$
$$+ e^{-a_q(t'+t)}\langle h_{-q}(t'=0)h_q(t=0)\rangle. \quad (A.13)$$

Using (A.12), the second moment of the capillary mode is:

$$\langle h_{-q}(t')h_q(t)\rangle = \frac{kT}{2b_q} e^{-a_q|t'-t|}. \quad (A.14)$$

The spatio-temporal time correlation function is defined as:

$$G(m,y,t) = \langle (h(m,y,t) - h(0,0,0))^2\rangle$$
$$= 2[\langle h(m,y,t)^2\rangle - \langle h(m,y,t)h(0,0,0)\rangle] \quad (A.15)$$

and can be written:

$$G(m,y,t) = \frac{kT}{N_x N_y} \sum_{q_x,q_y} \frac{1 - e^{-a_q|t|} e^{-i(q_x m + q_y y)}}{b_q}. \quad (A.16)$$

The above expression gives the required time–space correlation function for steps at thermal equilibrium. In a continuous approximation it can also be written:

$$G(t) = \langle (h_m(y,t) - h_m(y,0))^2\rangle$$
$$= \frac{kT}{\pi^2} \int_0^\pi \int_0^\pi \frac{1 - e^{-a_q|t|}}{b_q} \, dq_x \, dq_y. \quad (A.17)$$

### A.2. Analytical approximations

Various analytical expressions of $G(t)$ can be obtained depending on the considered time and the relative values of the parameters. For an anisotropic ($\eta_x \ll \eta_y$ and $\Gamma_x \ll \Gamma_y$) system in the rough phase ($V_{\text{loc}} = 0$), $a_q$ can be written under the form:

$$a_q = \frac{4\Gamma_y \eta_y}{kT}\left[\frac{\Gamma_x}{\Gamma_y}(1-\cos(q_x)) + (1-\cos(q_y))\right]$$
$$\times \left[\frac{\eta_x}{\eta_y}(1-\cos(q_x)) + (1-\cos(q_y))\right]. \quad (A.18)$$

Depending on the leading term within (A.18), one gets the following analytical expressions. For very short times ($t \ll (1/4\Gamma_y)(T/\eta_y)$), one can use the expansion:

$$1 - e^{a_q t} \approx \frac{4\Gamma_y \eta_y t}{kT}[1 - \cos(q_y)]^2, \quad (A.19)$$

that is $q_x$ independent and allows getting the linear time dependence:

$$G(0, 0, t) \approx 4\Gamma_y t. \quad (A.20)$$

For larger times, the term $(1 - e^{-a_q t})$ differs from 0 only in a narrow $q_y$ region close to 0 where $a_q t \approx 0$ and defined by $\cos(q_y) \approx \sqrt{1 - (kT/4\eta_y \Gamma_y t)}$. This is illustrated in Fig. 17 giving a graph within the $(q_x, q_y)$ plane of the argument of the sum within Eq. (11). Providing

$$\frac{1}{4\Gamma_y} \frac{kT}{\eta_y} < t < \frac{kT}{4\eta_y \Gamma_y} \left( \frac{\eta_x}{\eta_y} + \frac{\Gamma_x}{\Gamma_y} \right)^{-2}$$

one gets:

$$1 - e^{a_q t} \approx 1 - \exp\left(-\frac{4\Gamma_y \eta_y t}{kT}\right)[1 - \cos(q_y)]^2, \quad (A.21)$$

that is again $q_x$ nearly independent. Using $q_y$ expansion for $a_q$ and $b_q$ and the continuous approximation allows obtaining:

$$G(0, 0, t) = \frac{2kT}{\pi} \int_0^\pi \frac{1 - \exp\left(-\frac{\Gamma_y}{kT}\eta_y q_y^4 t\right)}{\eta_y q_y^2} \, dq_y$$
$$= 2\frac{\Gamma(3/4)}{\pi} \left( \frac{kT}{\eta_y} \right)^{3/4} (\Gamma_y t)^{1/4}. \quad (A.22)$$

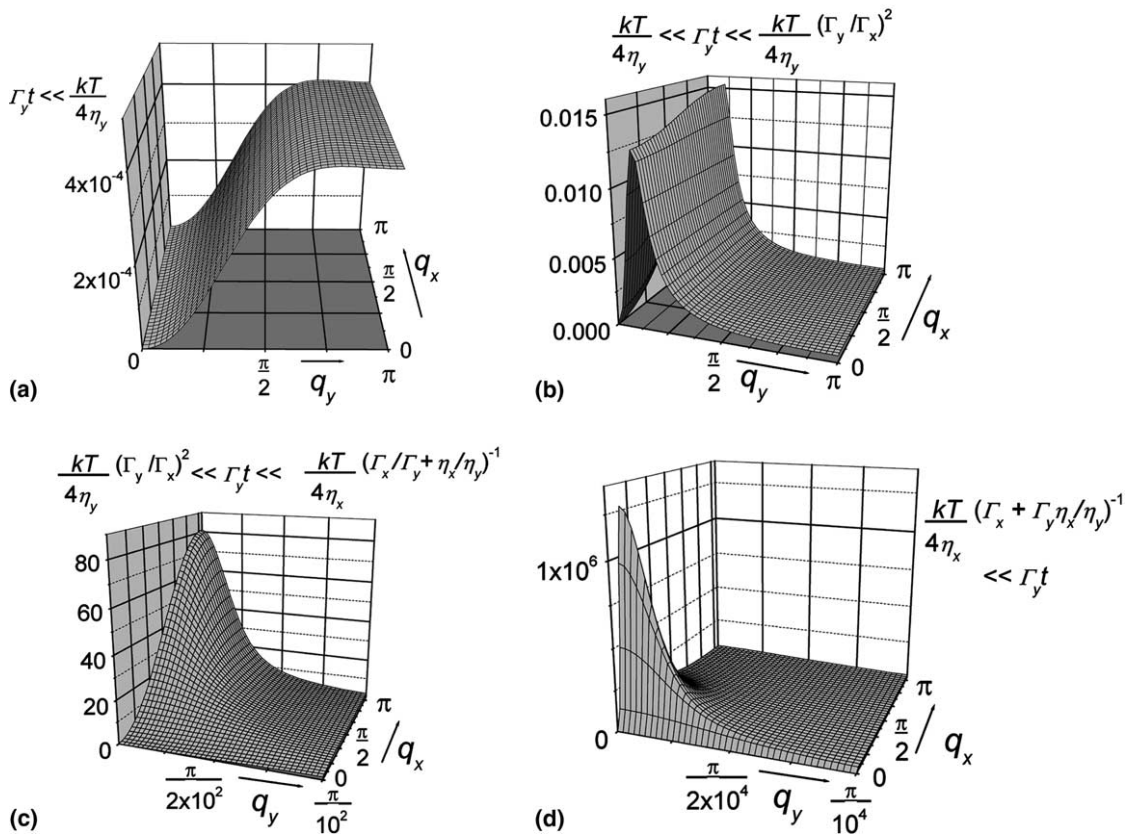

Fig. 17. Plots within the $(q_x, q_y)$ plane of the argument of the sum in Eq. (11): $(kT/N_x N_y)((1 - e^{-a_q t})/b_q)$. For various values of $\Gamma_y t$: (a) $10^{-3}$, (b) 1, (c) $10^4$ and (d) $10^8$. Parameters are: $\eta_x/\eta_y = 10^{-8}$, $\eta_y/T = 100$, $V_{loc} = 0$ K; $\Gamma_x = 1$ s$^{-1}$, and $\Gamma_x/\Gamma_y = 10^{-2}$.

For longer times so as

$$\frac{kT}{4\eta_y \Gamma_y}\left(\frac{\eta_x}{\eta_y}+\frac{\Gamma_x}{\Gamma_y}\right)^{-2} < t < \frac{kT}{4\eta_x \Gamma_y}\left(\frac{\eta_x}{\eta_y}+\frac{\Gamma_x}{\Gamma_y}\right)^{-1}$$

the $(1-e^{-a_q t})$ term within Eq. (11) vanishes everywhere excepted in a narrow region close to the $q_x$ axis (see Fig. 17) where $a_q t \approx 0$ and where the following relation holds:

$$\frac{\eta_x}{\eta_y}+\frac{\Gamma_x}{\Gamma_y} \gg 1-\cos(q_y). \qquad (A.23)$$

Providing $\eta_x/\eta_y \ll 1-\cos(q_y)$, $b_q$ remains only $q_y$ dependant and one has:

$$G(0,0,t) = \frac{2kT}{\pi^2 \eta_y}$$
$$\times \int_0^\pi \int_0^\pi \frac{1-\exp\left(-2\frac{\eta_y}{kT}\Gamma_y\left(\frac{\eta_x}{\eta_y}+\frac{\Gamma_x}{\Gamma_y}\right)(1-\cos(q_x))q_y^2 t\right)}{q_y^2}\,dq_x\,dq_y, \qquad (A.24)$$

where the very presence of $q_y^2$ instead of $q_y^4$ within the exponential term makes $G(t)$ vary as $t^{1/2}$:

$$G(0,0,t) = 8\pi^{-3/2}\left[\frac{kT}{\eta_y}\left(\frac{\eta_x}{\eta_y}+\frac{\Gamma_x}{\Gamma_y}\right)\Gamma_y t\right]^{1/2}. \qquad (A.25)$$

For very long times

$$t > \frac{kT}{4\eta_x \Gamma_y}\left(\frac{\eta_x}{\eta_y}+\frac{\Gamma_x}{\Gamma_y}\right)^{-1}$$

the relation $\eta_x/\eta_y \approx 1-\cos(q_y)$ holds and $b_q$ becomes $q_x$ and $q_y$ dependant. Assuming $(\eta_x/\eta_y \ll \Gamma_x/\Gamma_y)$, one gets by integration in polar coordinates:

$$G(0,0,t) = \frac{kT}{\pi\sqrt{\eta_x \eta_y}}\left[\frac{1}{4}\ln\left(\pi^4 \frac{\eta_x}{kT}\Gamma_x t\right) + \text{cte}\right], \qquad (A.26)$$

whereas for $(\eta_x/\eta_y \gg \Gamma_x/\Gamma_y)$, one has:

$$G(0,0,t) = \frac{kT}{\pi\sqrt{\eta_x \eta_y}}\left[\frac{1}{4}\ln\left(\pi^4 \frac{\eta_x^2}{\eta_y kT}\Gamma_y t\right) + \text{cte}\right], \qquad (A.27)$$

where the added constant is numerically found equal to 0.7326.

The above analytical expressions are only valid within some range of the parameters summarized in Table 1. Their use are limited for extended well defined $t$, $t^{1/4}$, $t^{1/2}$ or $\ln(t)$ regimes. Usually, and more importantly when a cross over from one regime to another is expected, it is easier and more accurate to use the general expression (A.17).